\documentstyle[preprint,aps,epsfig]{revtex}
\renewcommand{\baselinestretch}{1}
\begin{document}
\title{Interference due to Coherence Swapping}
\author{Arun Kumar Pati$^{(1)}$ and Marek Zukowski$^{(2)}$}
\address{$^{(1)}$ Theoretical Physics Division, 5th Floor, Central Complex,
BARC, Mumbai-400 085, India}
\address{$^{(2)}$ Instytut Fizyki Teoretycznej i Astrofizyki,
Uniwersytet Gdanski, PL-80-952 Gdansk, Poland}
\date{\today}

\maketitle

\tighten

\def\ra{\rangle}
\def\la{\langle}
\def\ver{\arrowvert}

\newcommand{\ket}[1]{\left | \, #1 \right \rangle}
\newcommand{\BEQ}{\begin{equation}}
\newcommand{\EEQ}{\end{equation}}
\newcommand{\BEQA}{\begin{eqnarray}}
\newcommand{\EEQA}{\end{eqnarray}}


\begin{abstract}
We propose a method  called ``coherence swapping'' which enables us to
create superposition of a particle in two distinct paths, which is fed with 
initially incoherent, independent radiation.
This phenomenon is also present for the charged particles, and can be used to
swap the effect of flux line due to the Aharonov-Bohm effect.
We propose an optical version of experimental set-up to test the
coherence swapping. The phenomenon, which is simpler than
entanglement swapping or teleportation, raises some fundamental questions
about the true nature of wave-particle duality, and also opens up the possibility
of studying the quantum erasure from a new angle.
\end{abstract}

\vskip .5cm

PACS NO: 03.65.Bz, 03.67.Hk\\

Keywords: Quantum interference, measurement, qubits, Aharonov-Bohm effect

email:apati@apsara.barc.ernet.in\\
Pramana- J of Phys. {\bf 56} (2001) 1.\\

\vskip 1cm


{\bf 1. Introduction:}\\

One of the basic mysteries in quantum theory is the phenomenon of interference
of a quantum particle if it has more than one alternatives to choose from 
\cite{rf}. The famous Young's double slit experiment can be demonstrated 
with single photon or electron, which vividly shows this interference effect
in nature. In fact, most of the physical phenomena in micro world can be attributed to 
the quantum interference. In recent years the quantum interference effects have been 
exploited to achieve parallelism in quantum computers, which cannot be performed with 
a classical computer \cite{qc,slb}. The basic unit in quantum computer ``qubit'' is 
nothing but a linear superposition of two distinct bits which is capable of showing interference 
effect. In quantum interference (first order) the important requirement is the coherence
of a quantum state, which usually we tend to associate with a particle if it 
has come from a single source and made to pass through a double slit or through
a suitable device such as a beam splitter (as in a Mach-Zehnder  interferometer).
But can we imagine a situation where we can observe interference between two
particles coming from independent sources and passing through two independent
double slits?

In this paper we propose a scheme called  ``coherence swapping'' (CS) which enables us to 
create superposition of a particle in two distinct paths, which is fed with  initially incoherent, 
independent radiation. This allows us to observe interference between two independent sources,
which can be  quite far apart.  In case of charged particles, the coherence swapping method can be
used to swap a flux line in Aharonov-Bohm setting. For example, the effect of flux line can be 
transferred from one set of interfering paths to another set of paths which may be far apart and 
even may not see the presence of vector potential. The coherence swapping will have application 
in quantum computer \cite{qc,slb} where one can create a qubit from independent outputs after 
some information processing inside two independent quantum computers.  For example, one 
can take one branch of the computational state from one computer and other branch from 
another computer and then by performing CS operation one can create a new computational 
state.

The process of ``coherence swapping'' is analogous in spirit to the recently proposed scheme 
``entanglement swapping'' \cite{zzhe,zzw}, but they are different. The difference is that 
entanglement swapping can be used to create an ``ebit'', whereas coherence swapping can be 
used to create a ``qubit''. The entanglement swapping has also been verified experimentally
 \cite{bet,bet1} using parametric down conversion sources. The coherence swapping provides 
us a means to have interference ``out of nothing" (or more precisely, selected out of a complete 
noise). The idea of creating correlation  for independent emissions goes back to the fundamental 
paper by Yurke and Stoler \cite{ys}. However, the scheme we propose has not been realized 
before. This raises some question as to what is the true nature of wave-particle duality of a 
quantum particle. We hope coherence swapping will open up possibility of studying this duality 
and in particular the quantum erasure problem \cite{sew,hv} from a new perspective.   \\

{\bf 2. Coherence swapping}\\

 Let us consider the interference phenomenon in a typical Mach-Zehnder interferometer. A 
particle comes from an source $S_1$ and falls on a 50-50 beam splitter (BS). After the action 
of BS the the particle is in a superposition of being in the outputs $a$ and $b$:

\begin{equation}
\ver \Psi \ra = \frac{1}{\sqrt 2} (a^{*}+b^{*})\ver 0 \ra,
\end{equation}
where we have used the convention that the particle creation operators in the state of being in 
the given beam, $a^{*}$ and $b^{*}$, bear the same name as the beam, and $\ver 0 \ra$ is the 
vacuum state. As the later effect will be due to the particle indistinguishability we assume that 
we deal with bosons (but similar effects are expected for fermions). The interference arises due 
to the fact that without some additional measurements towards establishing in which path the 
particle is, its interaction with the beamsplitter does not reveal this information. This can be 
seen by allowing these two paths to recombine and superpose on the exit beamsplitter (after
passing through a  phase shifter $PS_1$  in say the arm $a$) and then putting two detectors the 
exit ports. After passing through phase shifters the state becomes

\begin{equation}
\ver \Psi \ra = \frac{1}{\sqrt 2} (e^{i\phi}a^{*}+b^{*})\ver 0 \ra,
\end{equation}
After they are recombined at $BS_2$ the state becomes (we drop the over all phase)

\begin{equation}
(\cos \frac {\phi}{2}  a^{*} + 
i \sin \frac{\phi}{2}  b^{*})|0 \ra, 
\end{equation}
where we used the convention that the transmitted beams are are denoted by the same letter as 
the input ones. This shows that the detector $0$ clicks with probability $\sin^2(\frac{\phi}{2})$ 
and the detector $1$ clicks with probability $\cos^2(\frac{\phi}{2})$. This is  the interference 
phenomenon.
   
Let us consider two independent sources $S_1$ and $S_2$ each emitting one particle. Let 
particle 1 pass through a beam splitter $BS_1$ as in a Mach-Zehnder interferometer. It splits 
coherently to two paths $a$ and $b$. We can write the state of the particle after passing through 
beam splitter as given in (1). Now suppose the second particle coming from the source $2$ 
passes through another beam splitter $BS_2$. After passing through $BS_2$ the state of
particle $2$ can be similarly written as

\begin{equation}
\ver \Psi' \ra = \frac{1}{\sqrt 2} (c^{*}+d^{*})\ver 0 \ra,
\end{equation}
where $c,d$ have the same meaning for the 2nd particle. If we allow the paths to recombine then 
particle $2$ will show the interference as before. Let us modify the interferometer set-up
in a slightly different way. Instead of recombining path $a, b$ and $c,d$ let us recombine the 
particles in the path $b, c$ and $ a, d$ by introducing suitable beam splitters $BS_3$, $BS_4$ 
and mirrors $M_1$, $M_2$ as shown in the figure-1. Now the combined state of the system in 
the interferometer setup can be written as

\begin{equation}
\ver \Phi,\Phi' \ra = \frac{1}{2}(a^{*}+b^{*})(c^{*}+d^{*})|0\rangle
\end{equation}
The important observation is that if paths $b$ and $c$ lead to a 50-50
beamsplitter $BS_3$, the following unitary transformation links
the inputs and the outputs (denoted by the subscript $out$)
$$b_{out}= \frac{1}{\sqrt{2}}(b+c)$$ 
and
$$c_{out}= \frac{1}{\sqrt{2}}(b-c).$$
Therefore behind the beamsplitter $BS_3$ the state reads as
\begin{equation}
\frac{1}{2}(a^{*}+{1\over\sqrt{2}}(b^{*}_{out}+c^{*}_{out}))({1\over\sqrt{2}}
(b^{*}_{out}-c^{*}_{out})+d^{*})|0\rangle.
\end{equation}
This can be rewritten as
\begin{eqnarray}
&\frac{1}{2}(a^{*}d^{*}+b^{*}_{out}{1\over\sqrt{2}}(d^{*}+a^{*})
+c^{*}_{out}{1\over\sqrt{2}}(d^{*}-a^{*})&\\ \nonumber
&+{1\over2}(b^{*2}_{out}+c^{*2}_{out}))
|0\rangle.&\\
\end{eqnarray}

Therefore we see that if one registers a {\it single} photon in the $b$ output 
of the mixing beamsplitter, the other photon continues its propagation in 
the state $${1\over\sqrt{2}}(d^{*}-a^{*})|0\rangle,$$ i.e. it is in a {\it coherent} 
superposition of being in the beam $d$ and $a$, whereas if a photon is registered in the 
$c$ output, the other one continues in a superposition, which is orthogonal to the previous 
one namely $${1\over\sqrt{2}}(d^{*}+a^{*})|0\rangle.$$
This shows that the particles passing through the path $a$ and $d$ 
are now in a pure state conditioned on the detection result
behind the mixing beamsplitter. If say only events of single particle counts
at $b$ are selected the other particle can reveal interference phenomena
(if say we put a phase shifter into the beam $d$, and then superpose 
the beams $a$ and $d$ on another beamsplitter).
This is a method to swap coherence from the primary pairs of possible paths to another pair
paths. Such a process of {\em creating first order single particle interference involving 
two initially incoherent independent paths we call coherence swapping.}  The coherence
swapping creates a ``qubit'' out of complete random noise, which is the key result of the paper.  \\

{\bf 3. Coherence swapping and Aharonov-Bohm effect}\\

It is known that if a charged particle encircles a flux line in a Young's
double slit experimental set-up then we observe a shift in the interference
pattern even though there is no magnetic field present along the path
\cite{ab}. The field is present {\em only} inside the solenoid. But since the vector
potential is present along the path, the electron wavefunction is affected and
that gives rise to a phase shift in the interference pattern. In the standard
description of Aharonov-Bohm effect the electron wavefunction is split
coherently into $\Psi(r) = \Psi_1(r) + \Psi_2 (r)$. In the presence of
vector potential the wavefunction is modified to

\begin{eqnarray}
\Psi(r) =  e^{\frac{ie}{ \hbar c } \int_{\rm{path 1} } A. dx } \Psi_1(r) +
 e^{\frac{ie}{ \hbar c } \int_{\rm{path 2} } A. dx } \Psi_2 (r).
\end{eqnarray}
When they are recombined the interference pattern 

\begin{eqnarray}
I \sim  \ver \Psi_1(r) \ver^2+
\ver \Psi_2 (r) \ver^2 + 2{\rm Re}(\Psi_1(r)^* \Psi_2(r) ) \cos \Phi
\end{eqnarray}
depends on $\Phi = e/\hbar c\oint \vec{A}d\vec{x}$, which is the dimensionless  flux enclosed 
by the solenoid, which the electron has never seen. This is a non-local magnetostatic effect.

If in the interferometer of fig. 1 one has a magnetic flux in the internal 
area (bounded by the internal paths of the device), the interference due to the 
above described coherence swapping depends on the total flux passing thorough  this internal 
area. Therefore one can concentrate all flux in, say just a corner of the device
(e.g. close to the beamsplitter $BS_1$) and still due to the coherence  swapping 
observe interference at the other end of the device, behind
the beasmpitter $BS_4$  (of course conditioned on the events behind 
$BS_3$). We skip the calculational steps, but one can easily check that
the final interference after $BS_4$ will depend on the flux
enclosed by the solenoid. This can be called flux swapping. Thus,
flux swapping allows us to transfer the effect of flux line from one
set of interfering paths to another set of paths which have no common origin.
    
{\bf 4. Optical experiment to test Coherence swapping}\\

 In this section we propose an optical experimental set-up to test the 
coherence swapping (which can be done with current technology)

\subsection {Summary of physics of parametric down conversion}

If one shines a strong linearly polarised monochromatic laser
beam, or a quasi-monochromatic laser light pulse, on a suitably
cut and oriented birefringerent crystal endowed with a high
quadratic nonlinearity some pump photons spontaneously fission
into pairs of photons of lower frequency (for historical reasons
called signal and idler). The process is quasi elastic.
Thus the frequencies of pump photon, $\omega$, signal, $\omega_s$,
and idler, $\omega_i$, satisfy $\omega\approx\omega_s+\omega_i \label{ph1} $ 
(for the pulsed pump this relation still holds, however in this case the pulse
frequency is not precisely defined). Photons can be observed only
at so-called phase matched directions at which all emission processes within the full 
illuminated zone of the crystal interfere constructively:
${\bf k}_p\approx{\bf k}_s+{\bf k}_i, \label{ph2}$
i.e., the emissions are  strongly correlated directionally (again,
for the pulsed case ${\bf k}_p$ is not precisely defined). Due to
the dependence of the speed of light on frequency,  phase matching
within a crystal cannot occur for all frequencies, and all
emission directions, and thus into a given direction only specific frequencies are emitted.

Consider a pulsed pump. We assume that  the probability of a
multiple emission from a single PDC is low,  the laser pulse is
not too short, i.e., the nonmonochromaticity of the pulse will not
blur too much the strong angular correlation of the emissions (due
to the effective energy and momentum conservation within the
crystal). Thus, the photons can be still described as emitted in
specified, very well defined directions.

Under the  approximations that: (i) one has perfect phase matching
and only two phase matched directions are singled out; (ii) the
idler and signal frequencies satisfy perfect energy conservation
conditions with the pump photons, which is described by a sharply
peaked at the origin function $\Delta_L(\omega-\omega_i-\omega_s)$
which approaches Dirac's delta for $L\rightarrow\infty$ ($L$
symbolically represents the crystal's size); (iii) the pump pulse
is described as a classical wave packet (no-depletion) with one
single direction for all wave vectors (the frequency profile of
the pulse will be denoted by $V(\omega)$), the state of the photon
{\it pair} emerging from the PDC source (plus the filtering
system) via the beams $a$ and $d$ can be written as
\begin{eqnarray}
&|\psi_{ad}\rangle = &\nonumber\\ &\int d\omega_1 d\omega_2
d\omega_0 \Delta_L(\omega_0 - \omega_1
-\omega_2)g(\omega_0)&\nonumber\\ &\times
f_a(\omega_1)f_d(\omega_2) |\omega_1;a\rangle|\omega_2;d\rangle,
\\ \nonumber\label{4'}
\end{eqnarray}
where, e.g., the ket $|\omega; e\rangle$ describes a single photon
of frequency $\omega$ in the beam $e$, the function $g$ represents
the spectral content of the pulse,  $f_e$ is the transmission
function of the filtering system in the beam $e$ (a pinhole and/or
a filter).

The amplitude to detect a photon at time $t_{x'}$ by a detector
monitoring the beam $x'$ and another one at time $t_{y'}$ by a
counter in the beam $y'$, provided the initial photon state was,
say, $|\psi_{xy}\rangle$, can be written as
$A_{xy}(t_{x'},t_{y'})=(\langle t_{x'};x'|\langle
t_{y'};y'|)|\psi_{xy}\rangle$, where
$|t;b\rangle=\frac{1}{\sqrt{2\pi}} \int d\omega e^{i\omega
t}|\omega;b\rangle$ \cite{zzhe}. The elementary amplitudes of the detection
process  have  a simple, intuitively appealing, form \cite{zzw}
\begin{equation}
A_{xy}(t_{x},t_{y})=\frac{1}{\sqrt{2\pi}}\int dtG(t)F_x(t_{x}-t)
F_y(t_y-t), \label{amplitude}
\end{equation}
where the functions denoted by capitals are the Fourier
transforms: $H(t)=\frac{1}{\sqrt{2\pi}}\int d\omega e^{i\omega
t}h(\omega)$, for $h=f$ or $g$. The convolution of the filter
functions in (\ref{amplitude}) reveals one of the basic properties
of the PDC radiation: the time correlation between the detection
of the idler and the corresponding
 signal photon is entirely determined by the
bandwidth of the detection system. E.g. this implies that in the
limit of no filtering, when the functions $F(t)$ are approaching
$\delta(t)$, the time correlation extremely sharp (of the order of
femtoseconds), what can be illustrated by somewhat mathematically
incorrect limiting case of (\ref{amplitude}), namely $G(t_x)\delta(t_{x}-t_y),$
 (in reality, one also has to take into account the phase matching
function $\Delta$, and this imposes a sharp but finite time
correlation for the PDC process. However, just a
single filter will blur this correlation to around the inverse of the filter's bandwidth,
$\Delta T\approx 1/\Delta\nu$ (the coherence time of the filtered
radiation). The function $G(t)$ represents the temporal shape of
the laser pulse and its presence in the formula simply indicates
that (barring retardation) the signal and idler can be produced
only when the pulse is present in the crystal.

If the birefringent crystal is cut in such a way that the so
called type I phase matching condition is satisfied, both PDC
photons are of the same polarisation (if the pump beam is an
ordinary wave the down converted photons are extraordinary).  Due
to the phase matching condition (\ref{ph2}) (single) photons of
the same frequency are emitted into cones centred at the pump
beam. By picking photons from a specially chosen cone one can have
PDC radiation with both photons of equal frequency
$\frac{1}{2}\omega_p$. The selection can be done by a suitable
pinhole arrangement in a diaphragm behind the crystal and/or with
the use of filters.

\subsection{Proposed observation of coherence swapping}

We take two separate down conversion crystals, A, B, however
pumped by the same pulsed laser. The pump beam is beam-split in
such a way that the pulses enter both crystals at exactly the same
moment of time. We choose emission directions from the two
crystals in such a way that the emitted photons are degenerate in
frequency.

From time to time both sources emit spontaneously a pair of
photons each. We direct the idlers from each source to two trigger
detectors. The role of these trigger firings is the following one:
if both trigger detectors fire we know that the sources emitted
additionally one signal photon each. In front of both trigger
detectors one finds two identical filters. Their role is to the
define the coherence time of the detected radiation.

We place two 50-50 beam splitters, $BS_1$, $BS_2$, and direct to the
first one the signal of the source A, and to the second one the
signal of the source B. Each photon enters its beam splitter via
one input port. Of course, upon leaving the beam splitters, both
photons are in a coherent superposition of being in one or the
other exit beam of the beam splitter.

We direct one output of each beam splitter into one of the input
ports of yet another beam splitter, $BS_3$, behind which we place two
detectors. If only one of the detector fires, and this firing is
due to one photon only, then the other photon is in a coherent
superposition of being in one or the other exit beam (not fed into
$BS_4$) of the primary beam splitters $BS_1$, $BS_2$. If one wants to have
just one superposition state in beams $a$ and $d$, one should introduce a 
phase shift of $0$ or $\pi$ in one of the  beams, depending on which of
the detectors behind the beam splitter M fired (this can be thought to be 
the `classical communication' component here, if one seeks parallels of this 
experiment with the teleportation process).

The source must be pulsed in order to warrant that the parts of
wave  packets of the two photons that impinge on the BSM overlap
in the exit beams of it. Only in this way have the required total
erasure of the information: which of the sources contribute to the
click at behind $BS_3$. This warrants the required superposition
(coherence) to form in the other two beams.
One can test all this in an interference experiment on the beams.\\

The visibility of the obtained interference behind a device which
is composed of a phase shifter and a beamsplitter, $BM_4$, can be
obtained in the following simple way. One assumes that the filters in beams
leading to the trigger detectors (behind $BS_3$) are identical, and that the
functions have the following structure:
$F_f(t)=e^{-\frac{i}{2}\omega_p^ot}|F_f(t)|$,
$G(t)=e^{-i\omega_p^ot}|G(t)|$, where $\omega_p^o$ is the central
frequency of the pulse, then interference fringes in the joint
probability to have joint counts in  four detectors, that is the
two trigger detectors, one of the detectors behind BSM, and one of
the detectors behind BSX, behaves as $1-V\cos{(\phi)}$ with 
the visibility $V$  given by
\begin{eqnarray}
V=\frac {\int d^4t|A_{ad}(t_{a},t_d)A_{bc}(t_{b},t_{c})
A_{bd}(t_{b},t_d)A_{ac}(t_{a},t_{c}) |} {\int d^4
dt_|A_{ad}(t_{a},t_d)A_{bc}(t_{b},t_{c})|^2 },
\end{eqnarray} where $d^4t=dt_adt_bdt_cdt_d$.

If one specifies, for simplicity, all the functions as gaussians,
$exp{\left[-\frac{1}{2}(\omega-\Omega)^2/\sigma^2\right]}$, where
$\Omega$ is the mid frequency and $\sigma$ the width, the formula
for the visibility reads:
\begin{equation}
V=\left(\frac{\sigma^2_p}{\sigma^2_p +
\sigma^2_f}\right)^{\frac{1}{2}}.
\end{equation}
where $\sigma^2_p$ is the pulse frequency spread,  $ \sigma^2_f$
it the width of the filters.

The feasibility of this scheme has already been tested in the 
teleportation and entanglement swapping experiments \cite{bouw}, 
\cite{bet,bet1}. \\

{\bf 5. Conclusions}\\

 To conclude, we have proposed a new way of generating coherence between
 independent paths of two different quantum systems. The process of 
measurement,
 which was thought to be a hindrance to coherence, can be suitably designed
 to create coherence. Using coherence swapping method, one can swap the effect
 of flux line to a distant site. Finally, we
 have proposed an optical realisation of coherence swapping method. 
The phenomenon seems to be the simplest possible process involving 
interference of independent sources of quantum particles. Much simpler than 
entanglement swapping or teleportation, however it  shares
 with them the basic property:
possibility of a state preparation at a remote place
with the use of wave vector collapse and 
classical transfer of information (this latter feature is inherently
associated with the conditional nature of the interference
behind $BS_4$). One can generalise the concept of coherence swapping for
multi path and muliti particle interference set ups similar to the idea of entanglement
swapping for multiparticles \cite{bose}.
We hope that this will open up the possibility of studying
the wave particle duality and quantum erasure for independent particles.\\

Acknowledgement: AKP wishes to thank Sam Braunstein and P. van Loock for
discussions. He thanks EPSRC for  the financial support to carry out the
research. M\.Z was supported by the University of Gdansk Grant No
BW/5400-5-0032-0. \\

\renewcommand{\baselinestretch}{1}


\end{document}